\documentclass[aps, prl, twocolumn, floatfix, superscriptaddress, reprint, 10pt]{revtex4-1}  

\usepackage{bm}
\usepackage{graphicx}
\usepackage[usenames]{color}
\usepackage{microtype}
\usepackage{amsmath}
\usepackage{physics}
\usepackage{hyperref}

\begin{document}

\title{Learning the Electrostatic Response of the Electron Density through\\ a Symmetry-Adapted Vector Field Model}

\author{Mariana Rossi}
\affiliation{Max Planck Institute for the Structure and Dynamics of Matter, Luruper Chaussee 149, 22761 Hamburg, Germany}

\author{Kevin Rossi}
\affiliation{Department of Materials Science and Engineering, Delft University of Technology, 2628 CD, Delft, The Netherlands}
\affiliation{Climate Safety and Security Centre, TU Delft The Hague Campus, Delft University of Technology, 2594 AC, The Hague, The Netherlands}

\author{Alan M.~Lewis}
\affiliation{Department of Chemistry, University of York, Heslington, York, YO10 5DD, UK}

\author{Mathieu Salanne}
\affiliation{Institut Universitaire de France (IUF), F-75231 Paris, France}\affiliation{Physicochimie des \'Electrolytes et Nanosyst\`emes Interfaciaux, Sorbonne Universit\'e, CNRS, F-75005 Paris, France}

\author{Andrea Grisafi}
\email{andrea.grisafi@sorbonne-universite.fr}
\affiliation{Physicochimie des \'Electrolytes et Nanosyst\`emes Interfaciaux, Sorbonne Universit\'e, CNRS, F-75005 Paris, France}

\begin{abstract}

A current challenge in atomistic machine learning is that of efficiently predicting the response of the electron density under electric fields. We address this challenge with symmetry-adapted
kernel functions that are specifically derived to account for the rotational symmetry of a three-dimensional vector field. We demonstrate the equivariance of the method on a set of rotated water molecules and show its high efficiency with respect to number of training configurations and features for liquid water and naphthalene crystals. We conclude showcasing applications for relaxed configurations of gold nanoparticles, reproducing the scaling law of the electronic polarizability with size, up to systems with more than 2000 atoms. By deriving a natural extension to equivariant learning models of the electron density, our method provides an accurate and inexpensive strategy to predict the electrostatic response of molecules and materials.
\end{abstract}

\maketitle 

The electrostatic response of the electron density is a fundamental property of matter. It determines vibrational Raman and sum-frequency cross-sections~\cite{raim+19prm,hong+18njp}, and the optical dielectric constant of materials~\cite{sohier15jcp,cudazzo11prb}. Furthermore, it defines the induced electronic polarization of metallic systems~\cite{Dufils2019,Zhang2020}, of central interest for electrochemical energy storage~\cite{Jeanmairet2022} and electrostatic catalysis~\cite{Che2017,Che2018,Ke2022}. This electrostatic property is defined as the static linear response function of the electron density under the application of a homogeneous electric field along the three Cartesian directions; as such, it can be formally represented as a continuous three-dimensional vector field. 

The calculation of electronic density responses via density-functional theory (DFT) requires the evaluation of the first-order perturbative term in the density expansion around the ground-state, in absence of any external fields. This response can be obtained through suitable  finite-difference calculations. Density-functional perturbation theory (DFPT), however, provides a more elegant route to compute this quantity, allowing the treatment of both periodic and aperiodic systems on an equal footing~\cite{shang+njp2018,baroni2001,Gonze2024}. These calculations come with an added simulation cost involving a larger prefactor with respect to ground-state calculations, and a scaling with the number of electrons that is the same as that of DFT; this aspect vastly limits the system sizes that can be afforded, as well as the number of calculations that can be performed.

Bypassing the need for large computational resources is one of the central goals of machine-learning (ML) methodologies in atomistic simulations. In this context, ML models that are made equivariant under three-dimensional spatial symmetries have proved to be very efficient in learning electronic densities of molecules~\cite{gris+19acscs, fabr+20chimia,Cuevas2021,Jorgensen2022}, solids~\cite{lewis+21jctc,Achar2023}, liquids~\cite{Rackers2023,Grisafi2023}, and surfaces~\cite{Lv2023,grisafi2023prm,Grisafi2024,feng2024}. 
The equivariant nature of these methods drastically decreases the amount of data necessary to train reliable models, and typically enhances the overall accuracy by making predictions more robust to numerical noise. Nevertheless, building a fully equivariant framework for the prediction of vector-field properties represents a current challenge of atomistic machine learning. 
This problem has been tangentially faced in the description of spin degrees of freedom, as those entering the construction of ML interatomic potentials for magnetic materials~\cite{domina22prb,drautz20prb}. In that case, the atomistic representation must be able to capture the symmetries of the vector-field, but the learning target, i.e., the electronic energy, is invariant under rotations.

In this work, we address the challenge of achieving an equivariant atom-centered representation of the vector field and preserving the rotational symmetry in the prediction target, which is here the density response to applied electric fields. {From a ML model perspective, an advantage of directly targeting the density response rather than its derived properties (e.g., polarizabilities) is that the response is inherently local, making it more likely to yield robust predictions for highly polarizable systems.} Previous works by some of the authors have proposed methods to address this problem in the context of symmetry-adapted Gaussian process regression~\cite{Lewis2023, grisafi2023prm} (SA-GPR), by extending the SALTED method for predicting electron densities~\cite{lewis+21jctc,Grisafi2023}. These methods relied on a modification of the descriptor of the local atomic environment (features), in order to encode the information about the direction (and the strength) of the applied electric field. Recently, a similar approach has been adopted in the context of neural-network architectures~\cite{feng2024}, which also makes use of field-induced embedded atom density (FI-EAD) features~\cite{Zhang2019jpcl}. 
In Refs.~\citenum{grisafi2023prm} and~\citenum{feng2024}, these models were successfully applied to predict the density response of metallic slabs along the direction perpendicular to the surface, thus representing, effectively, the prediction of a scalar field. In Ref.~\citenum{Lewis2023}, the prediction target was the full vector field of the density response of molecules and insulating solids or liquids. However, the prediction of each Cartesian component was obtained through independent SA-GPR models. Learning and prediction were proven to be successful, but the rotational equivariance of the model was not formally obeyed. 

As we will see, an equivariant formulation of the learning problem capable of consistently taking into account the vector-field symmetries presents significant advantages over previous methods. 
We will show the superior accuracy of our approach in predicting the density response of benchmark datasets that include isolated water molecules, as well as supercells of liquid water and molecular crystals. Additionally, we will demonstrate applications to metallic nanoparticles of increasing size that display a nonlocal polarization, enabling the inexpensive calculation of polarizability tensors at large length scales.

In line with electron-density learning methods based on a linear atomic-orbital representation~\cite{gris+19acscs,lewis+21jctc,Rackers2023}, the response function of the electron density $n_e$ under an applied electric field $E_k$ can be written as a linear combination over a set of atom-centered basis functions $\phi_{n \lambda\mu}$. {Assuming a non-periodic system for simplicity, we can express  the continuous vector field at any given point~$\boldsymbol{r}$~as}

\begin{equation}\label{eq:response}
   \frac{\partial n_e(\boldsymbol{r})}{\partial E_k} = \sum_{i n \lambda\mu} c^k_{i n \lambda\mu} \phi_{n \lambda\mu}(\boldsymbol{r}-\boldsymbol{R}_i)\, ,
\end{equation}
where $k$, $n$ and $\lambda\mu$ are the Cartesian, radial and spherical harmonics indexes, respectively, and $\boldsymbol{R}_i$ are atomic positions. Reference quantum-chemistry calculations of this quantity can be  performed using a DFPT approach, as well as by finite differences of the electron density under a sufficiently small electric field along the three Cartesian directions. For periodic charge densities, these calculations need to be performed taking into account the modern theory of polarization~\cite{spal+12jssc}. The set of expansion coefficients, $c^k_{i n \lambda\mu}$, can then be extracted by using standard density-fitting techniques. Our goal is to derive a  kernel-based approximation of the coefficients $c^k_{i n \lambda\mu}$, by extending well-established SA-GPR methods for the prediction of $n_e$~\cite{gris+19acscs,lewis+21jctc,Grisafi2023}.

We start by noticing that the vectorial character of Eq.~\eqref{eq:response} can be equivalently expressed in terms of $\ket{1k}$ angular momentum states. This property allows us to represent the rotational symmetry of the expansion coefficients within the tensor product space $\ket{\lambda\mu, 1k} \equiv \ket{\lambda\mu}  \otimes \ket{1k}$. 
A similar scenario is encountered in the context of molecular Hamiltonians~\cite{Nigam2022}, where the matrix elements between atomic orbitals transform under rotations as the tensor product between pairs of angular momenta. In that case, one can take advantage of an orthogonalizion procedure to recast the learning target into a set of irreducible angular components that can be regressed independently of each other. Because of the non-orthogonal nature of the basis functions used for the expansion of Eq.~\eqref{eq:response}, however, this strategy cannot be effectively adopted for the present problem~\cite{Grisafi2023}. The rotational symmetry of the coefficients $c^k_{i n \lambda\mu}$ must be handled altogether in the learning model. 

From the previous discussion, the problem of providing an equivariant approximation of the $c^k_{i n \lambda\mu}$ can be addressed by deriving kernel functions that  express  the structural similarity between pairs of atomic environments $i$ and $j$, together with the geometric correlation between spherical tensors of order $\lambda \otimes 1$.  Following the rationale of Refs.~\citenum{glie+17prb,gris+18prl}, this can be achieved with $SO(3)$ integrals that include tensor products of Wigner $D$-matrices associated with the rotational symmetries we aim to enforce.
After some algebraic manipulations, detailed in the Supporting Information (SI), we find that kernel functions adapted to $\lambda\otimes 1$ symmetries can be written in terms of an irreducible angular momentum decomposition. Specifically, we obtain the central result:

\begin{equation}\begin{split}\label{eq:cgsum}
    K^{\lambda \otimes 1}_{\mu k,\, \mu'k'} = \sum^{\lambda+1}_{l=|\lambda-1|} &\braket{\lambda\mu,1k}{lm} \braket{\lambda\mu',1k'}{lm'}K^l_{m m'}\, ,
\end{split}\end{equation}
where $\braket{\lambda\mu,1k}{lm}$ are the Clebsch-Gordan coefficients used for the composition of angular momenta, such that $l=\{|\lambda-1|,\lambda,\lambda+1\}$ and $m=\mu+k$. Crucially, the irreducible components $\boldsymbol{K}^l$ in Eq.~\eqref{eq:cgsum} can be recognized as spherical tensor kernels of order $l$, like those originally derived in the prediction of  molecular tensors~\cite{gris+18prl}, and later used for the prediction of electron densities~\cite{gris+19acscs}. 
This makes the calculation of Eq.~\eqref{eq:cgsum} straightforward by accessing the relevant triplets of $\boldsymbol{K}^l$ for each value of $\lambda$ that enters the representation of Eq.~\eqref{eq:response}. Equivariance in $O(3)$ can be finally achieved by enforcing $\lambda \otimes 1$ inversion symmetry, which implies considering symmetric $\boldsymbol{K}^l$ for $l=|\lambda\pm 1|$ and antisymmetric $\boldsymbol{K}^l$ for $l=\lambda$.  

Following the latest formulation of the SALTED method~\cite{Grisafi2023}, the $\lambda\otimes 1$ kernels so derived can be used to perform equivariant predictions of the coefficients $c^k_{i n \lambda\mu}$ by recasting the learning problem into a low dimensional reproducing kernel Hilbert space (RKHS). 
This is done by first selecting a sparse set of $M$ atomic environments that best represent the structural variability in the dataset, following the subset of regressor approximation commonly used in GPR methods~\cite{quin+05jmlr}. 
Upon including this selection within the construction of $\boldsymbol{K}^{\lambda \otimes 1}$, we find that the dimensionality of the learning problem is increased by roughly a factor of 3 with respect to the case of the electron density, consistently with having incorporated into the model the vectorial symmetry of the density response. 
To limit the curse of dimensionality, we introduce in this work a Gaussian damping factor that decreases the value of $M$ for increasing values of angular momentum, i.e., $M_\lambda=M_0\, e^{-0.05\lambda^2}$, thus considerably reducing the final  RKHS size. A thorough discussion about the learning method reported in the SI.

\begin{figure}[t!]
    \centering
    \includegraphics[width=0.48\textwidth]{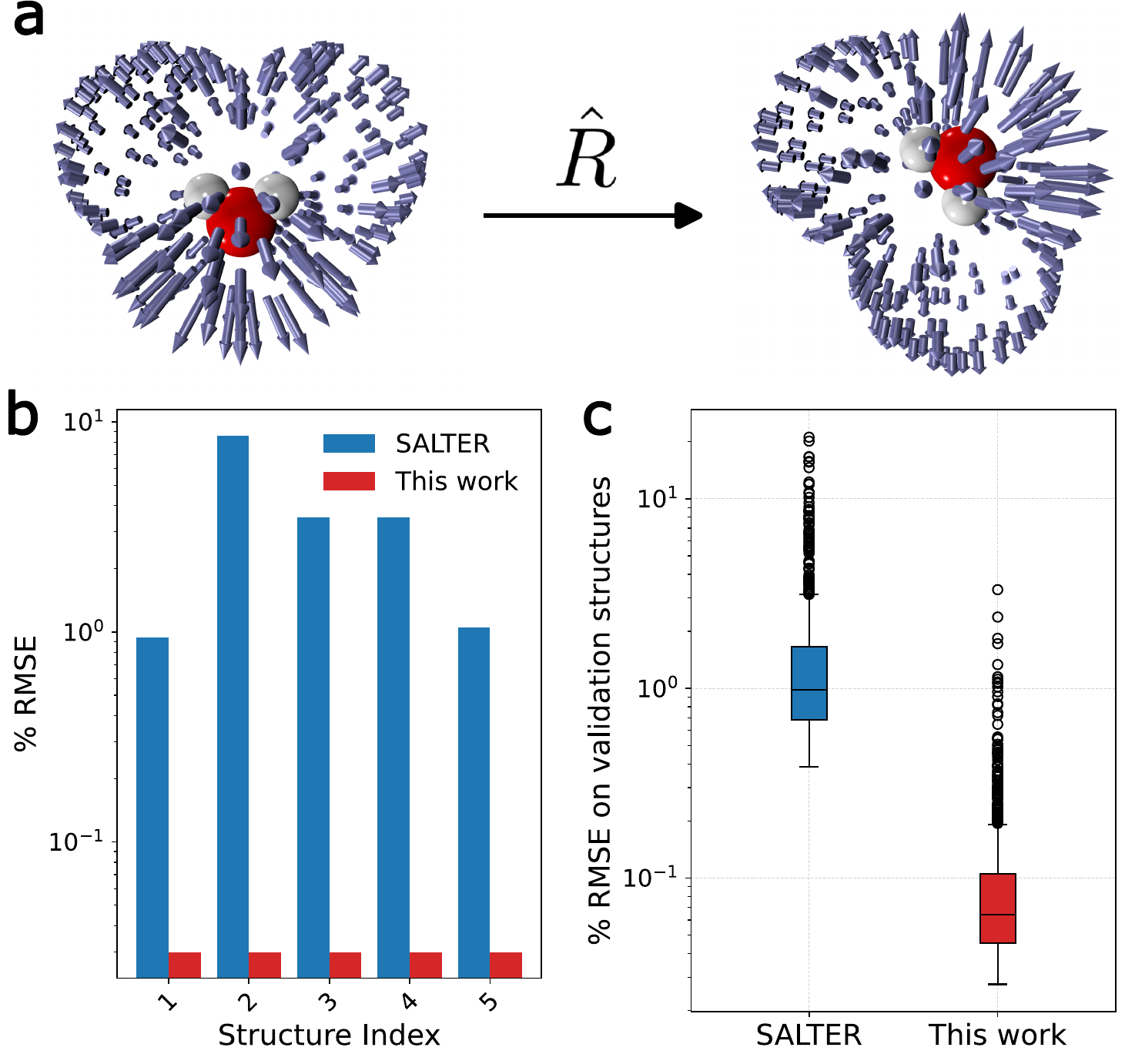}
    \caption{(a) Visualization of the vector field rotation corresponding to the electron density response of a water molecule to an applied homogeneous electric field. Vectors at a distance of 2~{\AA} from the closest atomic position are shown.  (b) Percentage root mean square errors (\% RMSE) of the density response prediction on five water conformers that are only rotated with respect to one another. The equivariant method of this work correctly yields the same error for these structures, while the method from Ref.~\citenum{Lewis2023} (SALTER) displays larger errors which are different for each conformer. (c) Box plot of all errors on the validation set of 900 structures, including rotated and distorted structures. While the average errors are small for both methods, the equivariant method of this work presents errors that are one order of magnitude smaller.}
    \label{fig:water-monomer}
\end{figure}

We test our method on a dataset of 1000 isolated water molecules presenting $100$ rigid configurations that are randomly rotated, plus $900$ that have a distorted geometry with a dipole moment aligned along the $z$-axis. This dataset has proven to be a useful benchmark for testing various equivariant methods~\cite{gris+18prl,Nigam2022}, as it allows us to simultaneously assess both the learning capability of the model and the effective incorporation of the desired rotational symmetries. Reference calculations of the density response function were obtained from DFPT, using the FHI-aims electronic-structure package.~\cite{blum+09cpc,shang+njp2018} An example of the computed vector field representing the density response between two mutually rotated water molecules is depicted in Fig.~\ref{fig:water-monomer}-a.

We train our model on a dataset of $N = 100$ randomly selected configurations -- thus leaving the remaining 900 for testing -- and compare the results with those that can be obtained using the SALTER method from Ref.~\citenum{Lewis2023}. As both methods are derived as an extension of SALTED, we are able to perform this comparison with completely equivalent parameters, most notably a rather small number of sparse environments ($M_0=100$). 
As a consequence of learning independent models for each Cartesian component of the density response, SALTER is not capable of automatically recognizing pairs of vector fields that are simply rotated with respect to each other.

This problem is well illustrated in Fig.~\ref{fig:water-monomer}-b by a comparison of the prediction errors reported for five water molecule configurations that are rigidly rotated with respect to one another. While SALTER displays inhomogeneous errors that fluctuate around 1\% RMSE, our method correctly produces entirely equivalent predictions that are almost 100 times more accurate. This highlights the effectiveness of symmetry adaptation in capturing the otherwise complex variations of the vectorial field under a three-dimensional rotation of the atomic structure.
For completeness, we report in Fig.~\ref{fig:water-monomer}-c a summary of the prediction errors of all the 900 test configurations, 90\% of which are aligned along the $\hat{z}$ direction. We observe a 50-fold improvement in prediction accuracy, when comparing SALTER with the fully equivariant approach of this work. We expect that if most of the molecules were not aligned, we would observe a more pronounced difference between the accuracy of both methods.

\begin{table}[h!]
    \centering
    \begin{tabular}{c|cc|cc}
         & \multicolumn{2}{c|}{Liquid Water} & \multicolumn{2}{c}{Naphthalene} \\
         Method $\backslash$ $N$ & 40 & 200 & 40 & 200 \\
         \hline
         SALTER & 13.22 & 11.64 & 4.72 & 4.57 \\
         \textbf{This work} & \textbf{9.26} & \textbf{8.87} & \textbf{2.86} & \textbf{2.78}
    \end{tabular}
    \caption{\% RMSE in the predicted electron density responses across a test set of 100 liquid water structures and 100 naphthalene crystal structures using both SALTER and the symmetry-adapted vector field method presented in this work. Results are shown for different training set sizes $N$, highlighting the data-efficiency obtained through symmetry adaptation against the more data-hungry SALTER method.}
    \label{tab:water_naphthalene}
\end{table}

We continue by comparing the performance of the equivariant approach and SALTER for more realistic systems, namely the liquid water and naphthalene datasets described in Ref.~\citenum{Lewis2023}. The liquid water dataset consists of cubic cells of side 9.67 \AA, containing 32 water molecules; the naphthalene structures are $2 \times 2 \times 1$ supercells of the P2$_1$/a naphthalene crystal, with 8 molecules per supercell. Each dataset contains 500 structures, of which 100 are reserved for validation. The \% RMSE in the electron density responses predicted by each method are summarized in Table \ref{tab:water_naphthalene}. Once again, we see a significant improvement upon moving to the fully equivariant method, which is clearly manifested by requiring a smaller amount of data to achieve a comparable prediction accuracy. In particular, we find that predictions that can be obtained by training our model on $N=40$ configurations are more accurate than those associated with training SALTER on $N=200$ configurations. This property is observed not only in the size of the training set, but also in the number of sparse atomic environments used to define the RKHS of the learning problem. In fact, while SALTER models were defined using a relatively large value of $M$, i.e., $M=3000$ for water and $M=2000$ for naphthalene for every value of $\lambda$, our symmetry-adapted method uses for both systems $M=500$ environments when $\lambda = 0$, falling to just $M=42$ environments when $\lambda = 4$. 
We conclude our discussion by applying the method on a dataset of gold nanoparticles of various sizes, reaching prediction targets that are challenging to calculate with full ab initio theory.
These systems are expected to display a large polarization effect that is related to a nonlocal charge rearrangement when approaching the metallic limit of a vanishing HOMO-LUMO gap~\cite{Ishida2020,Rajarshi2020}. For this reason, we include long-distance equivariant (LODE) features~\cite{gris-ceri19jcp} into the construction of each $\boldsymbol{K}^l$, a choice that has already proven successful to predict the electronic polarization of metallic slabs~\cite{grisafi2023prm}.
To investigate different size regimes, we generated a dataset of 255 relaxed nanoparticles, ranging from 28 to 139 atoms. 
We consider magic-size archetypes (icosahedra, decahedra, and fcc-truncated octahedra), as well as geometries obtained by randomly eliminating atoms from the cluster surface, and a successive geometry relaxation. Details about reference DFT calculations are reported in the SI. 

\begin{figure}[t!]
    \centering
    \includegraphics[width=8.5cm]{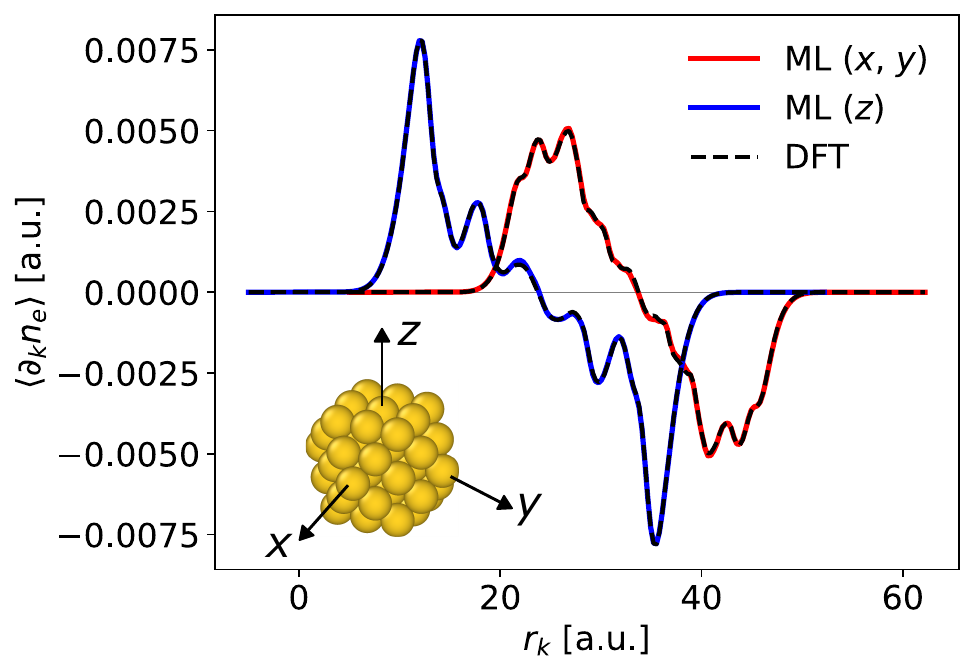}
    \caption{Comparison between the reference and predicted electron density response ($\partial_k n_e$), along a given Cartesian direction $k$, for a test nanoparticle made of 55 gold atoms. Response profiles along the three Cartesian directions are obtained by averaging the vectorial field components along the other two complementary Cartesian directions. Dashed black lines: DFT reference. Red and blue lines: ML predictions.}
    \label{fig:response-nano}
\end{figure}

We train our vector-field model on 200 randomly selected configurations and retain the remaining 55 for validation. The learning problem is then recast into a  low dimensional space of $M_0=200$ sparse atomic environments, following the Gaussian decay already described for $M_{\lambda}$ up to $\lambda=7$ spherical harmonics. A collective 15\% RMSE is measured over the entire test set, which we find to result in highly accurate predictions. 
As an example, we report in Figure \ref{fig:response-nano} the predicted density response in real space for a test 55 atoms cuboctahedron. We observe an excellent agreement with DFT by comparing the density response along the 3 Cartesian axes, reproducing the expected symmetry property of the selected structure along the $xy$-plane. 

\begin{figure*}[t!]
    \centering
    \includegraphics[width=16cm]{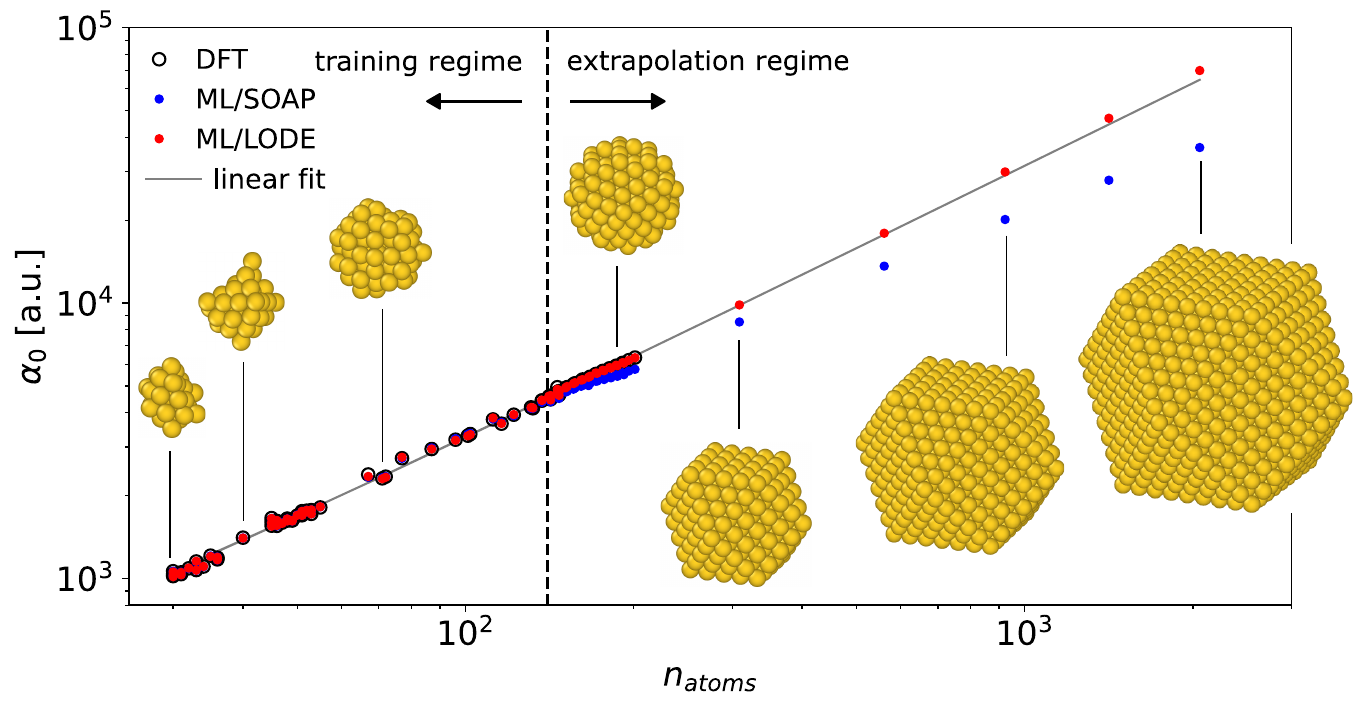}
    \caption{Isotropic components ($\alpha_0$) of the polarizability tensors of 90 gold nanoparticles of increasing size, as obtained from the corresponding electron density response coefficients. Empty circles: DFT reference. {Blue dots: ML prediction based on SOAP descriptors. Red dots: ML prediction based on LODE descriptors. Gray line: linear fit of $\alpha_0$ on the reference DFT values.} The vertical dashed line separates the domain of sizes used for training the vector-field model ($n_\text{atoms}<140$) from the large-scale domain of sizes used to validate the model under extrapolation conditions ($n_\text{atoms}>140$). No DFT reference is reported for the 5 largest  nanoparticles, which are chosen as unrelaxed cuboctahedron configurations made of 309, 561, 923, 1415, and 2057 atoms.}
    \label{fig:alpha-nano}
\end{figure*}

As a relevant application of the model previously trained, we consider the calculation of the electronic polarizability of the Au nanoparticles.
When considering isolated systems -- or systems presenting a distribution of electronic charge that vanishes before the cell periodic boundaries -- the polarizability tensor can in fact be directly computed as the first moment of the density-response function:
\begin{equation}\label{eq:alpha}
    \alpha_{kk'} = \int_{\mathcal{R}^3} d\boldsymbol{r}\, \boldsymbol{r}_k \frac{\partial n_e(\boldsymbol{r})}{\partial E_{k'}}  .
\end{equation}
Notably, analytical formulas can be derived that allows us to inexpensively compute $\alpha_{kk'}$ from the predicted $\lambda=0$ and $\lambda=1$ density-response coefficients; an explicit derivation is reported in the SI. We remark, however, that all  spherical harmonics components of the density response (up to $\lambda=7$ in this example) must still be considered during the training phase to account for the spatial overlap between basis functions.

An exemplary important question is that of assessing the capability of our model to reproduce the scaling law of the isotropic component of the polarizability tensor, defined as $\alpha_0=\frac{1}{3}(\alpha_{xx}+\alpha_{yy}+\alpha_{zz})$, with an increasing system size~\cite{Abhishek2024}. 
For this purpose, we perform reference DFT calculations for an additional set of 30 test relaxed configurations that include larger nanoparticles than those used for training the model, up to 201 atoms. These serve as benchmarks of our prediction accuracy in the extrapolative regime.
Five unrelaxed cuboctahedron geometries of 309, 561, 923, 1415, and 2057 atoms are finally included to the prediction targets, in order explore an asymptotically large size regime. For these, no reference value is computed because of the large computational cost.

Figure~\ref{fig:alpha-nano} reports prediction results for the whole set of 90 test structures considered. We find that our model can be used to accurately predict the expected value of $\alpha_0$ both within the training and extrapolation size domains with a comparable level of accuracy, achieving a collective error of only 1.2\% RMSE.  {In agreement with the classical electrodynamic response of ideal metallic spheres, $\alpha_0$ is expected to follow a cubic scaling law with the nanoparticle radius, which manifests as a linear increase of the polarizability with the number of gold atoms. At a fundamental level, this result derives from the localization of the density response at the nanoparticle surface when approaching the classical metallic limit~\cite{Snider1983}. A local model that relies on nearsighted SOAP descriptors cannot capture this collective surface effect and shows a sublinear scaling of $\alpha_0$ with respect to the number of gold atoms, as shown in Figure~\ref{fig:alpha-nano}. Instead, when including long-range information into the model through LODE structural features, the expected linear behaviour is recovered throughout the whole extrapolation set, which includes a tenfold increase in size with respect to the training geometries (see also Fig.~S1 in the SI).}  
To estimate the computational speedup obtained with our model, we parallelize the calculation on 192 AMD EPYC 9654 2.4 GHz CPU-cores. The observed prediction times are of 1.5 seconds for $n_\text{atoms}=309$ and of 11.6 seconds for $n_\text{atoms}=2057$. These timings represent a remarkable acceleration compared to DFPT or finite-field DFT calculations, estimated to be $>10^3$ using comparable computational settings.

Accurate results are also found in predicting the anisotropic part of the electronic polarizability. In order to obtain a single measure of this quantity, we compute the Frobenius norm of the traceless $\boldsymbol{\alpha}$-tensor. 
While most configurations present a relatively small anisotropic component, a few nanoparticles display a strong anisotropic response. 
We find that our method can effectively reproduce both limits, including, once again, those configurations that fall outside the regime of sizes used for training. A collective error of 11.9 \% RMSE on the anisotropic part is obtained throughout the test set. A more in-depth  discussion is reported in the SI (section VI). 

Following benchmarks ranging from molecular and periodic systems with diverse chemistry, to metallic clusters of increasing size that exhibit nonlocal polarizability,
we conclude that our equivariant vector-field approach offers an accurate and scalable framework for predicting the electrostatic linear response of the electron density in a cost-efficient manner. In particular, we confirm the vastly superior performance of this method with respect to similar approaches that do not incorporate equivariance with the learning target.
A key feature of our method is the streamlined incorporation of the vector-field rotational symmetry by a single Clebsch-Gordan sum over state-of-the-art spherical tensor kernels. We recognize that an iterative version of this procedure has been similarly adopted in the context of increasing the structural body-order within the construction of kernel-based ML potentials~\cite{Bigi2024}. Looking ahead, unifying these two approaches could enhance the capabilities of kernel methods in predicting generic electronic-structure properties within a shared mathematical framework. {Moreover, training on highly heterogeneous datasets will require the implementation of suitable strategies aimed at reducing the size of the learning problem~\cite{Lopanitsyna2023}.}

When it comes to the calculation of derived properties, the local nature of the density response allows us to perform highly transferable predictions of dipolar polarizability tensors via suitable integral procedures. While these tensors can directly enter the calculation of standard Raman signals, higher order polarizability moments could also be similarly computed to enable the calculation of surface-enhanced Raman spectra~\cite{Kedziora1999,Han2022}. 
Furthermore, having access to the  real-space distribution of the response function  could make it possible to treat spatially inhomogeneous induced polarizations, such as those encountered in tip-enhanced Raman scattering~\cite{Litman2023}. 
Beyond optical dielectric properties, the prediction of spatially inhomogeneous responses 
could also be used for tuning dielectric-consistent DFT functionals~\cite{Zhan2023}, relevant for improving the performance of DFT applied to semiconductor interfaces. 
{Finally, we envisage the application of our method to model the induced nonlocal polarization at electrified metal/electrolyte interfaces~\cite{Andersson2025}, thus enhancing the capabilities of first-principles atomistic simulations of electrochemical systems and electrocatalytic~processes.}

\section{Supporting Information}

Please see the Supporting Information for a thorough derivation of symmetry-adapted kernels, a discussion of the vector-field learning model, details about the dataset generation and ML parameters, the calculation of the polarizability tensor, complementary results.

Training and test geometries, together with the   DFT inputs used to generate the reference density-response data are free to download at~\url{https://zenodo.org/records/14678822}. An open source implementation of the density-response learning method is available at~\url{https://github.com/andreagrisafi/SALTED}, including example input files for training the model on the water monomer dataset.

\section{Acknowledgments}

This work was supported by the French National Research Agency under the France 2030 program (Grant ANR-22-PEBA-0002).  
This article is also based upon work from COST Action CA22154 - Data-driven Applications towards the Engineering of functional Materials: an Open Network (DAEMON) supported by COST
(European Cooperation in Science and Technology).
This work used the Dutch national e-infrastructure with the support of the SURF Cooperative using grant No. EINF-8985. The authors acknowledge HPC resources granted by GENCI, France (resources of CINES, Grant No. A0170910463).

\end{document}